# In Consideration of Indigenous Data Sovereignty: Data Mining as a Colonial Practice

Jennafer Shae Roberts and Laura N Montoya

Accel AI Institute, San Francisco, CA
info@accel.ai, Home page: http://www.accel.ai

**Abstract** Data mining reproduces colonialism, and Indigenous voices are being left out of the development of technology which relies on data, such as artificial intelligence. This research stresses the need for inclusion of Indigenous Data Sovereignty and centres on the importance of Indigenous rights over their own data. Inclusion is necessary in order to integrate Indigenous knowledge into the design, development, and implementation of data-reliant technology. To support this hypothesis and address the problem, the CARE Principles for Indigenous Data Governance (Collective Benefit, Authority to Control, Responsibility and Ethics) are applied. We cover how the colonial practices of data mining do not align with Indigenous convictions. The included case studies highlight connections to Indigenous rights in relation to the protection of data and environmental ecosystems, thus establishing how data governance can serve both the people and the Earth. By applying the CARE Principles to the issues that arise from data mining and neocolonialism, our goal is to provide a framework that can be used in technological development. The theory is that this could reflect outwards to promote data sovereignty generally, and create new relationships between people and data which are ethical as opposed to driven by speed and profit.

## 1    Introduction

Indigenous worldviews which hold that humans are a part of the land and cannot be separated from it have historically been usurped by the ideology that land is a commodity to be exploited for economic purposes. For example, access to resources such as water is globally justified to be controlled selectively and exclusively, as nature is increasingly commodified. Indigenous narratives stand in opposition to commodification[1][2] and need to be included, as now, we see this same commodification in data mining.

This paper argues that mining, whether of data or of the Earth, is not ethical. Mining metals from the earth is reflected in the term 'data mining'. On the surface, mineral mining appears to be a part of our modern world and how we extract metals necessary for many products that we use daily, such as smartphones and computers; however, upon closer examination, we can see the complexity and the harms that come to communities and the natural environment in this extraction process. Metal mining has negative implications for local Indigenous Peoples' health and the environment, a further implication of the colonial nature of the practice. Data mining is different from metal mining in several ways; one being that data is not something that occurs naturally, but that must be produced by individual people. Data mining is often compared to extracting resources, or as a sort of modern-day land grab,[3] another highly unethical colonial practice. Prioritising Indigenous Data Sovereignty (ID-SOV) would help to mitigate harm and provide guidance to not allow for disadvantageous data mining practices to take place.

The problem addressed by this paper is the lack of inclusion of Indigenous voices in the development of technology that relies on data, such as artificial intelligence, which therefore reproduces new forms of colonialism.

The research question addressed herein is how Indigenous Data Sovereignty can be integrated into the design, development, and implementation of data-reliant technology to ensure that Indigenous knowledge systems and practices are taken into account and that Indigenous communities are not left out of the process.

We cannot address colonialism without also addressing capitalism. Colonialism came first, and historical colonialism, with its violence and brutality, paved the way for capitalism.[4] We are at the dawn of a new stage of capitalism, following the path laid out by data colonialism, just as historical colonialism paved the way for industrial capitalism. We cannot yet imagine what this will look like, but we know that what lies at its core is the appropriation of human experience through the misuse of data.[5]

Not only is this a problem because it creates global inequality, capitalism notably threatens the natural environment. Its structural imperative is based on an insatiable appetite for growth and profit, causing overconsumption of Earth's material resources, not to mention overheating the planet.[6] For instance, mining cobalt in the Congo has detrimental effects not just on the earth, but on people's lives, utilising harsh child labour.[7] The Congo is where we get over 50% of the world's cobalt, an essential raw mineral found in cell phones, computers and electric vehicles, as well as in lithium batteries, in which we will see an increase in demand with the rise of renewable energy systems.[8] It is important to consider where hardware is sourced for building the technology that we use. Furthermore, not only is data mining causing harm to people and the environment in how it is being collected but also how it is being stored long-term. Data centres[9] alone account for 2% of human carbon emissions, rivalling that of the airline industry.[10] There are plans and efforts to lower emissions[11] from data centres, which needs to be done across industries, alongside efforts to address the underlying issues of dependence due to capitalism and consumerism.

> "What decolonial thinking in particular can help us grasp is that colonialism — whether in its historic or new form — can only be opposed effectively if it is attacked at its core: the underlying rationality that enables continuous appropriation to seem natural, necessary and somehow an enhancement of, not a violence to, human development.'' (Couldry and Mejias, 2019)[12]

There needs to be a major shift in societal ethics and away from continuous appropriation. A focus on decolonisation of data and the digital world is the primary focus for this work, and can radiate out to the physical world and how we affect it.

Data and data analytics have become increasingly important and interdependent in the digital age, especially with the rapid development of artificial intelligence (AI). Governments and other decision makers are heavily reliant on data for policy and decision making. As has been the case in much of our history, the unwilling targets of policy interventions are disproportionately Indigenous Peoples, whose enduring aspirations for self-determination over their own knowledge, information systems, institutions and resources has consistently been sabotaged by larger consolidated governments. Data is continually mined from Indigenous Peoples, without their input or permission on how their data is collected, used or applied.[13] However, as will be explored throughout this paper, ID-SOV and governments have been striving to change this and pull away from such a victimising narrative.

Shining light on various Indigenous perspectives, case studies are presented to decentralise the Global North agenda, with a focus on taking care of the natural world. This paper connects environmental concerns with ID-SOV, such as drawing parallels between data mining and mineral mining. We connect and expand on research on the impact of data colonialism and the decolonial turn in the digital sphere,[14] and how that presents in African indigenous communities, as well as in Mexico and Colombia.[15] The paper concludes that through applying the CARE principles (Collective Benefit, Authority to Control, Responsibility and Ethics) formulated by the International Indigenous Data Sovereignty Interest Group[16] to developing technologies which are based on data, we can build systems that value life over profit, therefore do the work to decolonise instead of reproducing neocolonialism.

There is a significant gap in the literature on the inclusion of Indigenous voices in the development of technology that relies on data, such as artificial intelligence. This gap is due to the lack of understanding of Indigenous knowledge systems and practices by non-Indigenous researchers and developers. The lack of inclusion of Indigenous voices in the development of data-reliant technology has led to a lack of sensitivity towards Indigenous communities and a lack of understanding of the issues faced by these communities.

This research paper aims to address this gap in the literature by exploring the importance of Indigenous Data Sovereignty in the design, development, and implementation of data-reliant technology. The paper connects the need for the CARE principles for Indigenous data governance to the unethical colonial practices of data mining and demonstrates these principles through reflections on case studies from the Global South. By doing so, this research paper provides a framework for integrating Indigenous knowledge systems and practices into the design, development, and implementation of data-reliant technology that is sensitive to Indigenous communities and promotes data sovereignty generally. That being said, this research is not meant to represent all Indigenous peoples, nor to speak for them. It is intended to put them at the centre of the conversation of decolonisation, where they have been so often left out in the past.

The paper aims to establish the justification and importance of the new initiative and articulate the research problem and objective, which are linked explicitly to the associated issues and problems. Specifically, the research question addressed by this paper is how Indigenous Data Sovereignty can be integrated into the design, development, and implementation of data-reliant technology to ensure that Indigenous knowledge systems and practices are taken into account and that Indigenous communities are not left out of the process. The paper begins by defining a selection of terms in section 2 that will aid in understanding the subject matter. Section 3 describes the objectives of the paper, followed by the methodology in section 4 which includes a literature review and details the CARE principles of Indigenous Data Sovereignty. Case studies on data mining in the African Indigenous context, as well as in Mexico and Colombia, are presented next in section 5. The discussion and limitations of this study can be found in section 6, and finally, the conclusions and recommendations are presented in section 7.

## 2    Definitions of Terms

In this section, we provide definitions and descriptions of the following terms which we use throughout this paper: Indigenous data sovereignty, indigeneity, decolonisation of data, neocolonialism, data colonialism and digital colonialism.

### 2.1    Indigenous Data Sovereignty (ID-SOV)

Before computers, even before the development of the written language, knowledge and data were continuously passed down from generation to generation by Indigenous Peoples, and yet ID-SOV is a relatively new concept, first being published in 2016.[17] ID-SOV has been defined as ". . . the right of Indigenous Peoples to own, control, access and possess data that derive from them, and which pertain to their members, knowledge systems, customs or territories."[18] In this paper we describe data, which is the fuel for many modern technologies including artificial intelligence, in a broad sense which considers cultural knowledge and heritage alongside personal information.

"Indigenous Peoples have always been 'data warriors'. Our ancient traditions recorded and protected information and knowledge through art, carving, song, chants and other practices."[19]

## 2.2 Indigeneity

To have the conversation about ID-SOV, it is imperative to discuss the challenge in defining what it means to be Indigenous. According to The United Nations Declaration on the Rights of Indigenous Peoples (UNDRIP), indigeneity has to do with first colonial contact, which can be quite difficult to determine in countries where colonists were not settlers. The term 'tribes', although useful, is problematic in its colonial origins. However, 'Indigenous' can encompass a wide range of ethnically diverse peoples, including tribes, such as the hill tribes in the Mekong River area of Southeast Asia.[20] A common element of Indigenous Peoples is a strong desire to maintain autonomy, while also resisting marginalisation and discrimination.[21] While working to decolonise, we must stress that the term 'Indigenous' was a separation created by colonists used to determine who was human, and who was less than human.[22] The fact that we still function from this foundation is inherently detrimental. In decolonising data in modern times, ID-SOV is the place to start.

## 2.3 Decolonisation of Data

If we centre on the rights of those who have been most marginalised by colonialism, methods by which we can continue the process of decolonisation will become progressively clearer.

To state ID-SOV as a human right is one thing; however, to see it carried out, we must detangle from a long history of manipulation of data on Indigenous peoples, who were historically demonised and dehumanised to justify settler colonialism. Today, in a world where neocolonialism is rife, we see how this has mutated to now appear as victimisation of Indigenous peoples. This narrative needs to change in order to empower Indigenous peoples in the digital age. According to the Global Indigenous Data Alliance (GIDA), building strategic relationships with global bodies and mechanisms is necessary to promote ID-SOV and governance internationally by providing a visible, collective approach.[23] These relationships and the adherence to ID-SOV principles will be of benefit to everyone, as well as the health of the planet.

## 2.4 Neocolonialism

Colonialism is a deeply rooted world system of power and control that plays out in ways that have become normal, however are highly unethical and harmful. In our modern world, which is so reliant on technology, colonialism and neocolonialism within data and the digital realm remain a fundamental problem. Neocolonialism was first defined in 1965 by Kwame Nkrumah, the first President of Ghana and founding member of the Non-Alignment Movement. Nkrumah said that neocolonialism was when the State outwardly has all the markers of international sovereignty, however, foreign capital is used for exploitation and imperialist powers have control. In this way, neocolonial investment increases rather than decreases the gap between wealthy and impoverished countries.[24] This is a more subtle yet dangerous form of colonialism that affects everyone around the world, and is being amplified in the digital sphere.

There is a strong separation between the dominant powers and the people and communities from which they profit. This is often perpetuated by viewing the Global North as separate from the Global South. Stefania Milan and Emiliano Treré presented a plural definition of the South(s) as place and proxy going beyond geopolitical denomination and embracing a multiplicity of interpretations, creating space for anywhere that people ". . . suffer discrimination and/or enact resistance to injustice and oppression and fight for better life conditions against the impending data capitalism."[25] In section 5 we cover case studies from the Global South to give examples of data mining and the CARE principles that exemplify the multitude of contexts.

## 2.5 Digital Colonialism and Data Colonialism

There are two ways neocolonialism is being discussed in socio-technical language: digital colonialism and data colonialism. These are parallel terms and may be considered one and the same, however we will focus here on how they have been described independently.

## 2.6 Digital Colonialism

When digital technology is used for social, economic, and political domination over other nations or territories, it is considered digital colonialism. Dominant powers have ownership over digital infrastructure and knowledge, which perpetuates a state of dependency within the hierarchy, situating Big Tech firms at the top and hosting an extremely unequal division of labour, which further defines digital colonialism.[26]

## 2.7 Data Colonialism

Data colonialism addresses Big Data in the context of the predatory practices of colonialism. Data as well as labour from the Global South unethically fuel the development of artificial intelligence and other data-run technology, unfairly quantifying and qualifying representations of humans, as well as humans themselves.[27] We see a pulley system of interdependence. However, the concentration of power is clear.

Data colonialism is sometimes seen as a subset of digital colonialism, as data colonialism concerns the abstraction of life into bits and bytes, whereas digital colonialism encompasses data but also the infrastructure and hardware that make up the digital world and the connection to the internet.[28]

# 3 Objectives

The purpose of this research is to outline a qualitative cross-cultural examination of existing literature on Indigenous Data Sovereignty and data decolonisation, alongside case studies which depict how to apply examples from the CARE principles for future reflection as an approach to disallow data mining practices in the development of technology. In section 4.2 below, we present the CARE principles in a simplified table which highlights each principle in accordance with its application in data ecosystems for future research. Our goal is to show how imperative it is to include the voices and perspectives of Indigenous peoples when discussing data mining, and to promote the decolonisation of technology.

The proposed methods in this research paper are compared with other similar approaches within a shared comparison framework that includes the CARE principles for Indigenous data governance. The CARE principles are connected to the unethical colonial practices of data mining, and the proposed methods are demonstrated through reflections on case studies from the Global South. The comparison framework includes an analysis of how the CARE principles can be applied to data-reliant technology in a way that is sensitive to Indigenous knowledge systems and practices and that ensures that Indigenous communities are not left out of the process. The proposed methods are evaluated based on their effectiveness in addressing the issues faced by Indigenous communities and their ability to promote data sovereignty generally.

# 4 Methodology

Our methodology includes the literature review below, followed by the CARE principles for Indigenous data governance. In this paper, we connect the need for the CARE principles to the unethical colonial practices of data mining, and demonstrate the CARE principles through reflections on case studies from the Global South.

## 4.1 Literature Review

This research is impacted by a few core sources which we connected together to reach our conclusions. The papers 'Data Colonialism: Rethinking Big Data's Relation to the Contemporary Subject'[29] and 'The Decolonial Turn in Data and Technology Research: What is at Stake and Where is it Heading?'[30] served to influence our decolonial approach, however they failed to highlight the CARE principles as a possible solution to data mining. The CARE principles were formulated by the International Indigenous Data Sovereignty Interest Group,[31] which is a network within the Research Data Alliance and is made up of individuals and nation-state based ID-SOV networks. However, this group does not directly address data mining, as we apply their principles in this paper. The case studies from Africa exemplify the harms of data mining on Indigenous communities, yet do not use the terms of the CARE principles. Two of the case studies we address are sourced from the book *Indigenous Data Sovereignty and Policy* (2021),[32] and they didn't include the CARE principles either, although we point out where they align with each of the principles in the Conclusions and Recommendations section.

Table 1. Literature Review

| Sources | Contributions | Missing |
|---|---|---|
| Kukutai and Taylor, *Indigenous Data Sovereignty* (2016) | Definition / origin of ID-SOV | Critique of data mining |
| Couldry and Mejias, *Data Colonialism* (2019), *The Decolonial Turn* (2021) | Decolonisation and critique of data mining | ID-SOV and CARE Principles |
| Carroll, Russo, Garba, Figueroa-Rodríguez, Holbrook, Lovett, Materechera, Parsons, et al. *The CARE Principles for Indigenous Data Governance* (2020) | CARE Principles | Critique of data mining |
| Rodriguez, *Indigenous Policy and Indigenous Data in Mexico* (2021) | Case study | Application of CARE Principles |
| Rojas-Páez, *Narratives on Indigenous Victimhood* (2021) | Case Study | Application of CARE Principles |
| Abebe, Aruleba, Birhane, Kingsley, Obaido, Remy, and Sadagopan. *Narratives and Counternarratives on Data Sharing in Africa* (2021) | Case Study: Data Mining | Application of CARE Principles |

As a reflection on the related work in Table 1, we juxtaposed the CARE Principles with the case studies and criticised data mining as an unethical practice. Data mining is in opposition to the CARE principles and ID-SOV, which need to be centred on to ensure adequate data rights for everyone. We focus on Indigenous peoples rights in order to avoid the reproduction of colonial practices and learn from the past by listening to the voices of those who have been the most colonised.

### 4.2 CARE Principles for Indigenous Data Governance

Here we examine the CARE principles of Indigenous data governance (Collective benefit, Authority to control, Responsibility, and Ethics) against the issues of data mining in an effort to move towards decolonising data. Open Data movements are concerning for ID-SOV networks due to the lack of protection for Indigenous Peoples. There is an increased push for greater data sharing, which can be seen in the widely-accepted FAIR principles (Findable, Accessible, Interoperable, Reusable).[33] However, these principles create tension in regards to how Indigenous Peoples' data is protected, shared and used. The FAIR principles are focused on how to make the technology more efficient, and ignore the fact that data comes from people and needs rights and protection. While the FAIR principles are data-centric and ignore the impact on ethical and socially responsible data usage, including power differentials and historic conditions considering the acquisition and usage of data, the CARE principles centre on the well-being of Indigenous Peoples and their data, and can be implemented alongside the FAIR Principles throughout data lifecycles to ensure collective benefit.[34] In order to encourage data collectors and users to be more aligned with Indigenous worldviews, the CARE Principles provide a framework for consideration of appropriate data use, as seen in Table 2.[35]

Table 2. CARE Principles for Appropriate Data Usage

| Data Mining Problem | CARE Principles[36] | Recommendations |
|---|---|---|
| Data Mining only benefits the miners and profiters | Collective Benefit | *Data ecosystems shall be designed and function in ways that enable Indigenous Peoples to derive benefit from the data* |
| Indigenous people have no agency over their resources or data | Authority to Control | *Indigenous Peoples' rights and interests in their own data must be recognised and their authority to control such data must be empowered. Indigenous data governance enables Indigenous Peoples and governing bodies to determine how they, as well as their lands, territories, resources, knowledges and geographical indicators, are represented and identified within data* |
| There is no system in place to protect Indigenous data rights | Responsibility | *Those working with Indigenous data have a responsibility to share how those data are used to support Indigenous Peoples' self determination and collective benefit. Accountability requires meaningful and openly available evidence of these efforts and the benefits accruing to Indigenous Peoples* |
| Data mining practices are unethical | Ethics | *Indigenous Peoples' rights and wellbeing should be the primary concern at all stages of the data life cycle and across the data ecosystem* |

The CARE principles for Indigenous data governance are unique in that they draw from, integrate, and build on the work of mainstream stakeholders focused on data for reuse (e.g., FAIR Principles) and the efforts of Indigenous-led networks and coalitions focused on Indigenous data governance and research control. The CARE Principles detail that the use of Indigenous data should result in tangible benefits for Indigenous collectives through inclusive development and innovation, improved decision-making, and increased capacity building. The CARE Principles also emphasise the importance of respecting Indigenous knowledge systems and practices and ensuring that Indigenous communities are not left out of the process.

Compared to other methods and models in similar fields, the CARE Principles for Indigenous data governance are unique in that they provide a framework for integrating Indigenous knowledge systems and practices into the design, development, and implementation of data-reliant technology that is sensitive to Indigenous communities and promotes data sovereignty generally. By doing so, the CARE Principles provide a basis for future research in this area and contribute to a growing body of literature on Indigenous Data Sovereignty.

## 5 Relating Case Studies to Indigenous Data Sovereignty and CARE Principles

Indigenous Data Sovereignty (ID-SOV) is not only relevant but necessary for creating fairer governance and a more prosperous future for Indigenous peoples.[37] This is exemplified and stressed in the following case studies, where we tie in connections to the CARE principles.

### 5.1 Data Mining in the African Context

Data colonialism can be used to describe some of the challenges of data mining, which reflect the historical and present-day colonial practices[38] such as in the African and Indigenous context. When we use terms such as 'mining' to discuss how data is collected from people, the question remains: Who benefits from this data collection? [39]

The aggregation and use of data can paradoxically be harmful to communities from which it is collected. Establishing trust is challenging due to the historical actions taken by data collectors while mining data from Indigenous populations. We must address the entrenched legacies of power disparities concerning what challenges they present for modern data sharing.[40] As of 2021, the Open Government Partnership (OGP) listed fourteen members and ten states which had enacted data protection in Africa, including: Burkina Faso, Cabo Verde, Côte d'Ivoire, Ghana, Kenya, Liberia, Malawi, Morocco, Nigeria, Senegal, Seychelles, Sierra Leone, South Africa, and Tunisia.[41] The OPG is a charter which aims to liberate government-controlled data and focuses on the principles of transparency, accountability and participation.[42] Table 3 demonstrates how the OPG principles and their recommendations align with the CARE principles.

**Table 3.** OGP Principles and Recommendations and Related CARE Principles

| Thematic Area in African OGP Data Protection Principles | Recommendations | Aligned CARE Principles |
|---|---|---|
| Transparency | The Right to Notification, Data Processing Registers | Authority to Control |
| Accountability | The Power to Investigate, The Power to Sanction, Regular Reporting | Responsibility and Ethics |
| Participation | The Right to Access Personal Data, The Right to Request the Correction or Deletion of Personal Data, The Right to Request the Correction or Deletion of Personal Data | Collective Benefit, Authority to Control, and Ethics |

### 5.2 Data Mining Case Study

One problematic example of data mining is occasions where non-government organisations (NGOs) attempt to 'solve' issues for marginalised groups, yet can inadvertently cause more harm than good.[43] For example, one European NGO attempted to address the problem of access to potable water in Burundi, while testing new water accessibility technology and online monitoring of resources which used data mining.[44]

The NGO failed to understand the community's perspective on the actual central issues and the potential harms of their actions. By sharing data publicly, which included geographic locations, the NGO put the community at risk. Collective privacy was violated and there was a loss of trust. Therefore, the CARE principles were violated, particularly *Collective benefit* and *Responsibility*. Note that Burundi is not on the above list from the OPG, and does not yet have data protection laws in place, nor a definition of personal data under Burundi law, however companies are required to gain consent prior to transferring personal data to third parties under some sectoral laws.[45] Western thought centres privacy as primarily a personal concern, however protection of collective identity also stands in a position of great importance for many African and Indigenous communities.[46][47] This example exhibits that trust cannot be formed on the foundation of power imbalances, which oppose each of the CARE principles. There is no *Collective benefit* if outside parties are in control, which disallows for the *Authority to control* data. *Responsibility* and *Ethics* are also missing in this study. Unforeseen and irreparable harm can be done to the wellbeing of communities when there is a lack of forethought into the ethical use of data, during and after the project. This creates a hostile environment upon which to build relationships of respect and trust.[48] This exemplifies neocolonialism in action: creating systems of victimisation and dependence which ultimately cause harm to the people it proposes to help.[49]

To conclude this case study, we can pose the ethical question: Is data sharing actually beneficial? Referencing the CARE Principles, local communities must be the primary beneficiaries of responsible data sharing practices. It is important to have specificity and transparency around who benefits from data sharing, and to make sure that it is not doing any further harm to the people behind the data.[50]

Heretofore, neocolonialism, ID-SOV, the CARE principles and data mining have been connected. In the next section, the connection to how we steer global data governance towards protecting the natural environment will be explored, examining case studies on metal mining in Mexico and protections against exploitation of resources and cultural data in Colombia.

### 5.3  Mexico: the Example of Mineral Mining and ID-SOV

Mineral mining in Mexico is one extractive process which profoundly impacts Indigenous communities and has only been promoted by recent Mexican presidents. In the last twelve years, 7% of Indigenous territories have been lost for the sake of mining alone, frequently failing to inform local Indigenous communities.[51][52] Without the knowledge or permission of local Indigenous peoples, external actors have historically conducted research to better understand the values of natural resources in Indigenous territories, demonstrating a lack of understanding of the implications of exploiting things such as minerals, timber, wildlife, plants, and water for the people who live there, in terms of health and environmental consequences, infrastructure, and investments.[53] As discussed in the introduction, the extractive process of mining data is a continuation of mining natural resources as a form of colonialism. The government has stepped up in recent times to protect Indigenous rights to their own culture and data, as well as the rights of Afro-Mexican peoples, supporting the *Authority to Control* principle, highlighted in the CARE principles.

Further relating to the *Responsibility* CARE principle, new laws have recently been enacted to protect Indigenous communities and Afro-Mexican communities and their heritage. January 2022 saw a vote by the Mexican Congress to approve the Federal Law for the Protection of the Cultural Heritage of Indigenous and Afro-Mexican Peoples and Communities.[54] This law includes protecting Indigenous and Afro-Mexican communities and their rights to property and collective intellectual property, traditional knowledge and cultural expressions, including cultural heritage, in an ". . . attempt to harmonise national legislation with international legal instruments on the matter, trying to give a seal of 'inclusivity' to minorities."[55]

Cultural heritage relates directly to data about communities, which is where the CARE principle of 'Collective Benefit' comes into play. The definition of intangible cultural heritage aligns with how we think of collective data and includes "uses, representations, expressions, knowledge, and techniques; together with the instruments, objects, artefacts, and cultural spaces that are inherent to them; recognized by communities, groups, and, in some cases, individuals as an integral part of their cultural heritage."[56] Cultural heritage represents data about a collective. By protecting cultural heritage, the lands and natural resources of Indigenous communities are also protected.

### 5.4  ID-SOV in Colombia

In Colombia, Indigenous inspectors have been appointed to monitor natural resources on reservations since 1987, which, if also applied to Indigenous data, would exemplify the CARE principles of *Responsibility* and *Ethics*. In 1991, Colombia approved their new Constitution recognising Indigenous rights, including ethnic and cultural diversity, languages, communal lands, archaeological treasures, parks and reservations, which they have traditionally occupied; and adopted programs to manage, preserve, replace and exploit their own natural resources.

[57]

The Colombian government's efforts and commitments to strengthen the dialogue on human rights have been recognised by political figures of the European Union. Patricia Llombart, Colombia's European Union Ambassador, stated that Colombia has shared values with the EU and is seen as a reliable and stable partner. Where the EU has been involved, international agreements, which include protecting Indigenous rights as well as labour rights and rights for children, have been signed in Andean countries.[58]

Further recognition of the protection of Indigenous data can be seen as an enactment of the CARE principles and ID-SOV in Colombia. "The protection of personal data is a constitutional and fundamental right in Colombia,"[59] stated Carolina Pardo, partner in the corporate department of Baker McKenzie in Colombia. Her article, *Colombia Data Protection Overview in DataGuidance*, referenced the Congress of Colombia enacted Statutory Law 1581 of 2012, which Issues General Provisions for the Protection of Personal Data ('the Data Protection Law'), which "develop the constitutional right of all persons to know, update, and rectify information that has been collected on them in databases or files, and other liberties and constitutional rights referred to in Article 15 of the Political Constitution."[60] This stands for all four of the CARE principles, namely *Authority to control* and *Collective benefit*.

# 6      Discussion

Data does not promote change automatically nor does it address issues of marginalisation, colonialism or discrimination. Additionally, there is little to no consideration given to combatting imbalances of power in negotiations and consultations led by major world governments.[61]

Even in recent times, sensitive COVID-19 data was mined and reused without consent from Indigenous Americans by the media, researchers and non-governmental organisations.[62] This was carried out under the assumption that publicising Indigenous communities' sensitive data would be helpful, when it actually caused unintentional harm in the process. This is a reflection of historical colonialism. Settler colonialists thought that they were 'helping' too, via ethnic cleansing and direct violence.[63] Tracing the histories can help us understand how to move towards decolonising data for the benefit of all.

This research paper is significant because it builds on previous work by connecting the need for the CARE principles for Indigenous data governance to the unethical colonial practices of data mining and demonstrating these principles through reflections on case studies from the Global South. The paper also highlights the limitations of previous work in this area, such as the failure to highlight the CARE principles as a possible solution to data mining and the lack of direct address of data mining by the International Indigenous Data Sovereignty Interest Group. By doing so, this research paper provides a framework for integrating Indigenous knowledge systems and practices into the design, development, and implementation of data-reliant technology that is sensitive to Indigenous communities and promotes data sovereignty generally.

Limitations of this study include the limitations of applying the CARE Principles for Indigenous Data Governance in technological development. There are further limitations of integrating Indigenous knowledge into the design, development, and implementation of data-reliant technology. For example, some limitations could be related to the lack of access to data or resources by Indigenous communities or the lack of understanding by non-Indigenous developers of Indigenous knowledge systems. Another limitation of the research paper is that the researchers are non-Indigenous and may not have a full understanding of Indigenous knowledge systems and practices. This is a limitation because the researchers may not have the same level of understanding of Indigenous knowledge systems and practices as Indigenous researchers would have. This could lead to a lack of understanding of the issues faced by Indigenous communities and a lack of sensitivity in the design, development, and implementation of data-reliant technology.

# 7 Conclusions and Recommendations

ID-SOV has a place as the laws continue to change to protect Indigenous rights. ID-SOV could help ". . . fill the gap regarding the lack of evaluations as an appropriate approach in the design and implementation of monitoring, evaluation, and learning (MEL) local systems, controlled and used by Indigenous communities."[64] (Rodriguez, 2021) MEL is a known framework for providing best practices and strategic tools for assessing the effectiveness of processes used by companies, governments, and NGOs. Rodriguez went on to list recommendations from the Organisation for Economic Co-operation and Development (OECD) to move forward on these issues.

> **The OECD recommends four main areas to strengthen Indigenous economies:**
> *1. improving Indigenous statistics and data governance*
> *2. creating an enabling environment for Indigenous entrepreneurship and small business development at regional and local levels*
> *3. improving the Indigenous land tenure system to facilitate opportunities for economic development*
> *4. adapting policies and governance to implement a place-based approach to economic development that improves policy coherence and empowers Indigenous communities[65][66]*

The recommendations to solve the problems caused by data mining according to the CARE principles include the following: First, data about Indigenous Peoples should benefit those that it belongs to, not hurt them while profiting third parties. This should be true for all people and all data, however, the current reality is quite different. Second, Indigenous Peoples should have the right to self determination of the data about them, to have authority on how they are represented. Third, there is a great responsibility in those dealing with Indigenous data to ensure self determination and benefit for the collective. Last but not least, ethical practices are incredibly important at all stages of data collection and application. If these principles can be integrated into systems of open data and utilised to inform data governance locally and globally, it could work towards promoting decolonisation and a more balanced relationship with data.

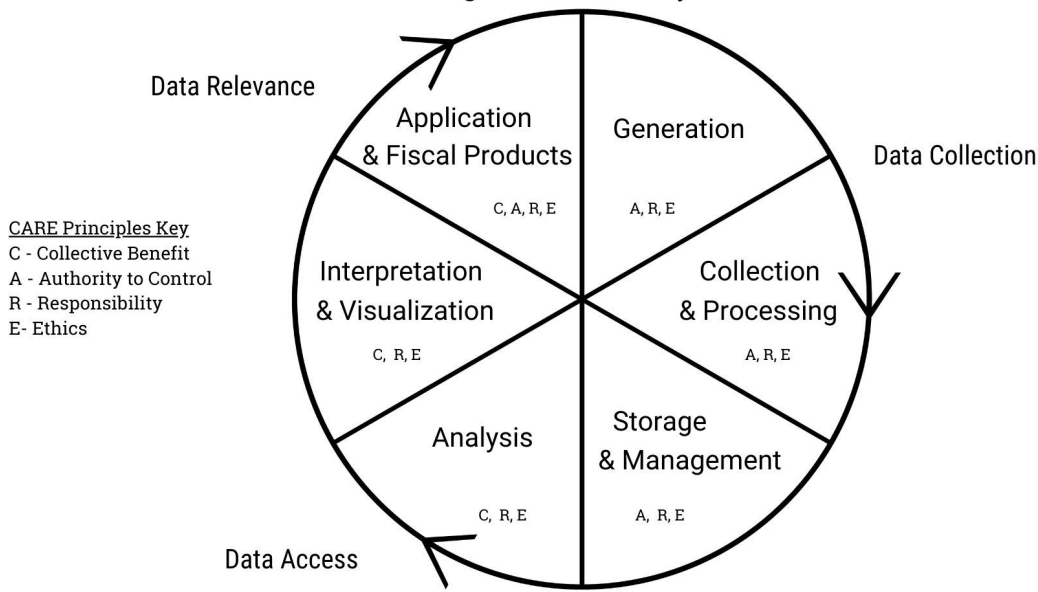

**Figure 1.** Data Lifecycle

CARE Principles Key
C - Collective Benefit
A - Authority to Control
R - Responsibility
E - Ethics

Through exploring the case studies from Africa, Mexico and Colombia, it is clear that the CARE principles of ID-SOV could fill the gaps in considering public policies for data governance for Indigenous peoples. There is a need to remediate three main data challenges: data collection, data access, and relevance in order to allow for access, use and control of Indigenous peoples' own data and information.[67] This is demonstrated in Figure 1, where we show where in the data lifecycle each of the CARE principles can and should be applied.

This is something that must be understood for data governance around the world. It is vital to note that there are varying local concerns in different regions, although all have been negatively influenced and impacted by long-standing exploitative colonial practices. It is imperative that we continue to educate ourselves and question broader narratives that stem from colonial roots. This is an overview that is far from exhaustive, but gives fresh perspectives by stressing the need to prioritise the voices of Indigenous Peoples as equally sovereign leadership.